\documentstyle[aps,preprint]{revtex}
\tightenlines
\begin{document}

\pagestyle{plain}
\setcounter{page}{1}
\setcounter{footnote}{00}

\renewcommand{\thefootnote}{\alph{footnote}}

\baselineskip=18pt 
\def\doublespaced{\baselineskip=\normalbaselineskip\multiply
    \baselineskip by 150\divide\baselineskip by 100}


%
%
\begin{titlepage}
\baselineskip=0.2in
\begin{flushright}
TUIMP-TH-98/98\\
\end{flushright}
\vspace{0.2in}
\begin{center}
{\large 
   Determining Top Quark CP Violating Dipole Couplings        
                from $e^+e^-\to t\bar t$   }\\
\vspace{.2in}
Hong-Yi Zhou\\
\vspace{.2in}
        Institut f\"{u}r Theoretische  Physik, Universit\"{a}t Heidelberg,
        Philosophenweg 16, D-69120 Heidelberg, Germany
{\footnote{Present mailing address}}\\
              and \\
    Institute of Modern Physics and Department of Physics,\\
    Tsinghua University, Beijing 100084, P.R. China\\

\end{center}
\vspace{.3in}

\begin{center}\begin{minipage}{5in}
\baselineskip=0.25in
\begin{center} Abstract\end{center}
We show how to determine the electric and weak dipole moments 
of the top quark simultaneously and independently 
from $e^+e^-\to t\bar t$
at $\sqrt{s}=500$ GeV NLC. To obtain the best 
accuracies with which the dipole moments can be measured, 
we apply the optimal observables to extract the CP violating 
effects and consider only purely hadronic, hadronic-leptonic
final state events. Results with left- and right- handed longitudinal 
polarized as well as unpolarized electron beams are given. 
We find that with $50fb^{-1}$ integrated luminosity, 
the dipole moments can be measured  to the 
accuracy of  $10^{-18}~e~cm$.   

\end{minipage}\end{center}

\vspace{.5in}


\end{titlepage}
\newpage

\baselineskip=18pt 
\renewcommand{\thefootnote}{\arabic{footnote}}
\setcounter{footnote}{0}

\section{ Introduction }
\indent
The top quark has its unique advantages in search for 
new physics beyond the Standard Model(SM) due to its 
large mass and decay properties. Many non-SM models 
predict several orders of magnitude 
large electric dipole moment(EDM) and weak dipole 
moment(WDM) of the top quark than the SM. These 
dipole moments give rise to CP violating effects 
in top quark pair productions. 
CP violating effects in  top quark pair production  
at $e^+e^-$ colliders have been widely studied\cite{review}-
\cite{recent}. Many of these studies concentrate  on the 
observables which use purely leptonic or hadronic-leptonic 
(including $bl_1^+\nu_{l_1}\bar b q_2\bar q'_2$ and 
$b\bar q_1 q_1'\bar b l_2^-{\bar\nu}_{l_2}$) 
final state events. Therefore, these studies can not give the 
best limits one can obtain from the experiments. To 
achieve the best goal, one needs to make use of the purely 
hadronic events which consist of 10/17 of the total events 
and to apply the optimal approach\cite{atwood92}
\cite{opt1}-\cite{opt3}. The best limits on the top quark 
dipole moments given in Ref.\cite{atwood92} are $\sim 10^{-19}~e~cm$.  
Our work has some similarities with  Ref.\cite{atwood92}. 
New features of our work are:(1) the W-boson polarization is not 
and indeed  can not be 
determined completely by a single event, but statistically from its 
decay products; (2) we consider how to measure the electric and weak 
dipole moments simultaneously and independently ; 
(3) we give more realistic estimation on the limits 
by taking into account decay branching ratios, luminosity and detection 
efficiency. In Ref.\cite{bern96}, the optimal observables 
are also used, but without using the spin information of the 
hadronically decayed top quark. We consider $\gamma t\bar t$ and 
$Z t\bar t$ including EDM and WDM couplings, and for simplicity, 
assume the top quark decays via SM interactions.

\section{ Calculations and Optimal Observables }
\indent
We assume the couplings of electron with $\gamma$ and $Z$ bosons 
take the standard model values:
\begin{eqnarray}
& &-ieg_e^V\gamma^\mu(1+\alpha_e^V\gamma_5),
\end{eqnarray}
where $V=\gamma,~Z$ and 
\begin{eqnarray}
& & g_e^\gamma=-1,~\alpha_e^\gamma=0,\\
& & g_e^Z=\frac{4\sin^2\theta_W-1}{4\sin\theta_W \cos\theta_W}~,
\alpha_e^Z=\frac{1}{4\sin^2\theta_W-1}~. 
\end{eqnarray}

The couplings between the top quark and $\gamma$, $Z$ bosons 
take the form:  
\begin{eqnarray}
& &-ie[g_t^V\gamma^\mu(1+\alpha_t^V\gamma_5)+(p_t-p_{\bar t})^\mu
(-id_t^V/e)\gamma_5],
\end{eqnarray}
where $p_t,~p_{\bar t}$ are the momenta of the top quark and top 
antiquark. $d_t^V$ is the dipole moment which we assume to have 
imaginary parts as well as real parts. 
We denote $\hat{d}_t^V=d_t^V/e$.  The other couplings are:
\begin{eqnarray}
& & g_t^\gamma=2/3,~\alpha_t^\gamma=0,\\
& & g_t^Z=\frac{1-\frac{8}{3}\sin^2\theta_W}{4\sin\theta_W \cos\theta_W}~,
\alpha_t^Z=-\frac{1}{1-\frac{8}{3}\sin^2\theta_W}~.
\end{eqnarray}
The standard model amplitude of $e^+e^-\to t\bar t$ is 
\begin{eqnarray}
& & M_0=ie^2\sum\limits_{V=\gamma,Z}g_e^Vg_t^V\bar v(p_{e^+})\gamma^\mu
(1+\alpha_e^V\gamma_5)u(p_{e^-})\bar u(p_t)\gamma_\mu(1+
\alpha_t^V\gamma_5)v(p_{\bar t})/(s-m_V^2), 
\end{eqnarray}
and the correction from the dipole moments is
\begin{eqnarray}
& &\delta M=ie^2\sum\limits_{V=\gamma,Z}g_e^V\bar v(p_{e^+})\gamma^\mu
(1+\alpha_e^V\gamma_5)u(p_{e^-})\bar u(p_t)[(p_t-p_{\bar t})_\mu(-i
\hat{d}_t^V)\gamma_5]v(p_{\bar t})/(s-m_V^2). 
\end{eqnarray} 
We shall assume the dipole moments are small enough that their 
quadratic contributions to the total cross section are negligible. 
Therefore the dipole moments contribute only to the CP violating 
effects through $2Re(M_0\delta M^\dagger)$ which is linear in 
$\hat{d}_t^V$.  To observe the CP violating effects, one needs 
to know the spins of the top quarks which can be determined 
statistically from their decay products. 
We assume the SM decay of the top quark and apply the narrow width 
approximations of the top quark and W-boson propagators:
\begin{eqnarray}                  
\frac{1}{|q^2_X-m_X^2+im_X\Gamma_X|^2}\rightarrow 
\frac{\pi}{m_X\Gamma_X}\delta(q_X^2-m_X^2)~,
\end{eqnarray} 
where $X$ stands for top quark and W-boson, $\Gamma_X$ is the width of 
$X$.  

The  cross section for reaction 
$e^+e^- \to t\bar t\to bl_1^+\nu_{l_1}\bar b l_2^-{\bar\nu}_{l_2}$ 
($b\bar q_1 q_1' \bar b q_2\bar q'_2$) can be written as
\begin{eqnarray} 
\label{cross}
d{\sigma}&=& \frac{\beta}{(8\pi)^{10}
s}\frac{\lambda_t |M_D|^2}
{m_t^2m_W^2\Gamma_t^2\Gamma_W^2}d\Omega_t d\Omega_{W^+}' d\Omega_{W^-}'
d\Omega_{l_1^+}'d\Omega_{l_2^-}' ~,   
\end{eqnarray} 
where $\beta=\sqrt{1-4m_t^2/s}$ and 
\begin{eqnarray} 
\lambda_t=(1-\frac{(m_W+m_b)^2}
{m_t^2})(1-\frac{(m_W-m_b)^2}{m_t^2})\approx (m_t^2-m_W^2)^2/m_t^4~,
\end{eqnarray} 
$d\Omega_{W^+}' (d\Omega_{W^-}')$ is the solid angle element of $W^+(W^-)$
in the rest frame of the (anti) top quark,  
$d\Omega_{l_1^+}'(d\Omega_{l_2^-}')$ denotes the     
solid angle element of $l_1^+(l_2^-)$ in the rest frame of $W^+(W^-)$, 
$|M_D|^2$ is the amplitude  square excluding the top quark and W-boson 
propagators after the decays of the top quarks:  
\begin{eqnarray}  
|M_D|^2 &=&|M_0|^2+2Re(M_0\delta M^{\dagger})~.
\end{eqnarray} 
If the electron(positron) beam is not polarized, additional 
spin average factor and summation are needed. 
In our calculations, $|M_D|^2$ is easily 
obtained from the amplitude of $e^+ e^-\to t\bar t$ by the following 
substitutions:
\begin{eqnarray} 
& & \bar u(p_t)\rightarrow \frac{g^2}{8}\bar u_b \gamma_\mu(1-\gamma_5)
(\rlap/p_t+m_t)\bar u_{\nu_1}\gamma^\mu(1-\gamma_5) v_{l_1}~,\\\nonumber
& & v(p_{\bar t})\rightarrow \frac{g^2}{8}\bar u_{l_2} \gamma_\mu(1-\gamma_5)
v_{\nu_2}(\rlap/p_{\bar t}-m_t)\gamma^\mu(1-\gamma_5) v_{\bar b}~,
\end{eqnarray} 
where $g$ is the weak $SU(2)$ coupling constant. The above expresssions  
are calculated numerically. 

Denoting $g_1=Re(\hat{d}_t^\gamma)$, $g_2=Re(\hat{d}_t^Z)$, 
$g_3=Im(\hat{d}_t^\gamma)$ and $g_4=Im(\hat{d}_t^Z)$, we can write 
 the amplitude square $|M_D|^2$ as 
\begin{eqnarray}
|M_D|^2 &=& \Sigma_0+g_1\Sigma_1+g_2\Sigma_2+g_3\Sigma_3+ 
g_4\Sigma_4~,   
\end{eqnarray} 
where $\Sigma_0=|M_0|^2$. For unpolarized beams, $\Sigma_{1,2,3,4}$ 
are independent. But for left- or right-handed polarized electron beam,
there are only two independent terms:
\begin{eqnarray}
\label{pol}
|M_D|^2_{L} &=& \Sigma_{0L}+g_1^L\Sigma_1^L+g_2^L\Sigma_2^L~,\\\nonumber
|M_D|^2_{R} &=& \Sigma_{0R}+g_1^R\Sigma_1^R+g_2^R\Sigma_2^R~,
\end{eqnarray} 
where $L,R$ stand for left- or right-handed polarized electron beam, and 
\begin{eqnarray}
\label{coe1}
& & g_{1}^L=g_{1}+(\xi-\eta) g_{2},~~~ g_{2}^L=g_{3}+(\xi-\eta) g_{4},
\\\nonumber 
& & g_{1}^R=g_{1}+(\xi+\eta) g_{2},~~~ g_{2}^R=g_{3}+(\xi+\eta) g_{4},
\end{eqnarray} 
where 
\begin{eqnarray} 
\label{coe2}
\xi&=&\frac{1-4\sin^2\theta_W}{4\sin\theta_W\cos\theta_W}\frac{s}
{s-m_Z^2},\\\nonumber 
\eta&=&-\frac{1}{4\sin\theta_W\cos\theta_W}\frac{s}{s-m_Z^2}.
\end{eqnarray} 
In our calculation, we have set the electron 
masses to be zero and do not consider the radiative corrections 
to $e^+e^-\gamma(Z)$ couplings. Therefore, even with  
polarized electron beams, only the initial CP eigenstates couple 
to $\gamma$ and $Z$.

To measure $g_i$, one needs to extract the CP violating effects 
which can be picked out by CP-odd observables. It has been 
shown in the literature\cite{atwood92}
\cite{opt1}-\cite{opt3} that the optimal observables defined 
below have the smallest statistical errors. 
The optimized CP-odd observables in the full final state 
phase space with unpolarized beams are defined by 
\begin{eqnarray}
O_{1i}=\frac{\Sigma_i}{\Sigma_0}~. 
\end{eqnarray} 
It is shown in Ref.\cite{opt2} that a linear transformation  
of the  above set of observables $O_{1i}$ are still optimal.

When the top quark decays hadronically, we can not distinguish 
quark and antiquark jet. For hadronic-leptonic 
events, the missing neutrino momenta can be fully reconstructed 
using energy momentum conservation equations, so that we are 
left with two fold ambiguity of the jet momenta. For purely 
hadronic events, we have four fold ambiguity.     
Considering this ambiguity, one can 
define alternatively the optimal observables:
\begin{eqnarray}
O_{2i}=\frac{\sum\limits_{j}\Sigma_i}{\sum\limits_{j}\Sigma_0},~~~ 
O_{4i}=\frac{\sum\limits_{j'}\Sigma_i}{\sum\limits_{j'}\Sigma_0}, 
\end{eqnarray} 
where the sum  $j$ is over the two possible assignments of the 
jet momenta to the quark and antiquark in hadronic-leptonic events.
$j'$ is over the possible assignments of the 
jet momenta to the quark and antiquark in purely hadronic events.  

All the above definitions can be applied to the polarized beam 
cases. We now consider separately  the polarized and unpolarized 
beams. 

We first look at the unpolarized beams. 
In this case, we can define four optimized observables for
each of the two kinds of final state events mentioned above. 
They can be separated into two categories: $O_{n1},O_{n2}(n=2,4)$ 
are $\hat{T}$-odd with $\hat{T}$ being the transformation that inverse 
the particle spins and momenta but does not interchange initial and 
final states, $O_{n3},O_{n4}$ are $\hat{T}$-even.  

The mean value of the observable $O_{2i}$ is defined as 
\begin{eqnarray} 
\langle O_{2i}\rangle&=&
\frac{\int d\sigma^+ O_{2i}^++   
\int d\sigma^- O_{2i}^-}{\int d\sigma^++\int d\sigma^-}~,   
\end{eqnarray} 
where the superscript $+,-$ mean that the integrations are over 
$bl_1^+\nu_{l_1}\bar b q_2\bar q'_2$ and 
$b\bar q_1 q_1'\bar b l_2^-{\bar\nu}_{l_2}$ 
final states, respectively.  
The mean value of the observable $O_{4i}$ is simply  
\begin{eqnarray} 
\langle O_{4i}\rangle&=&\frac{\int d\sigma O_{4i}}{\int d\sigma}~.   
\end{eqnarray} 

We can express the mean value of an observable $O_{ni}$ by
\begin{eqnarray}
\langle O_{ni}\rangle&=&c_{ij}g_j, 
\end{eqnarray} 
where $c_{ij}=\langle O_{ni}O_{nj}\rangle$. The symmetric matrix $c$  
is just the covariance of the observables.
Due to the different $\hat{T}$ symmetry, the matrix $c$ will 
be block diagonal between $i,j=1,2$ and $i,j=3,4$. Non-zero off-diagonal 
elements mean the statistcal dependences of the observables and 
therefore the couplings. It will be difficult to estimate  
the statistical error of a particular coupling without 
assuming the other couplings are zero.  
The way to solve the problem is to find a linear combination 
of the couplings and the observables that are statistically 
independent\cite{opt2}. To do so, one needs to diagonize the matrix 
$c$  by a matrix $A$. For simplicity, we choose $A$ to be 
orthogonal: 
\begin{eqnarray}
& &c'=A^{-1}cA,\\
& & O_{ni}'=A_{ji}O_{nj},\\
& & g'_i=A_{ji}g_j. 
\end{eqnarray}  
The 1$\sigma$ statistical error of coupling $g_i'$ is 
now given by 
\begin{eqnarray}
\label{err}
\Delta g_i'=\frac{1}{\sqrt{Nc_{ii}'}}, 
\end{eqnarray}
where $N$ is the number of events. To reduce the statistcal errors, 
one can combine the measurements of $O_{2i}'$ and $O_{4i}'$ to get 
a combined error $\Delta g_{ci}'$\cite{data}:
\begin{eqnarray}
\frac{1}{(\Delta g_{ci}')^2}=\frac{1}{(\Delta g_{2i}')^2}+
\frac{1}{(\Delta g_{4i}')^2}, 
\end{eqnarray}
where $\Delta g_{2i}'$ and $\Delta g_{4i}'$ are the errors of the 
measurements using $O_{2i}'$ and $O_{4i}'$,respectively.

From Eq.(\ref{pol}), we see that when the electron is left- or right-
handed polarized, we can only define two independent observables 
$O_{n1}, ~O_{n2}(n=2,4)$ which 
have different $\hat{T}$ parity. They are sensitive to 
$g_1^{L,R}$, $g_2^{L,R}$, respectively. Therefore we have a 
diagonal matrix $c$.   Only a linear combination 
of $d_t^\gamma$ and $d_t^Z$ can be measured with a particular 
polarization beam. The combination depends on the energy(cf.
(\ref{coe1})(\ref{coe2})).  

\section{Results and Conclusions}

We present our results of the matrix $c$ at $\sqrt{s}=500$ GeV 
unpolarized $e^+e^-$ collider in Table I. It shows the relations 
between the mean values of the observables and the coupling 
constants. The corresponding 
orthogonal matrix $A$ and the diagonal matrix $c'$ are given 
in Table II. Matrix $A$ is useful for extracting $g_i$ from $g_i'$ and 
for the calculation of $O_{ni}'$. In Table III., we give the results 
of the matrix elements $c_{ii}$ at the  collider with  polarized  
electron beams. 

We  assume: (1) the overall detection efficiency 
is $\epsilon=0.1$; (2) the integrated luminosity is ${\cal L}=50fb^{-1}$;
(3) the branching ratio of hadronic-leptonic 
final state events is $B_{lj}=0.29$($l=e,\mu$),the branching ratio of 
the purely hadronic events is $B_{jj}=0.46$. The number of events is given by 
\begin{eqnarray}
N&=&\epsilon{\cal L}\sigma B,
\end{eqnarray}
where $\sigma$ is the total $t\bar t$ production cross section, 
$B=B_{lj}$ or $ B_{jj}$. With $\alpha_{em}=1/128.8$ and  $m_t=176$ GeV, 
we get the following total cross sections for different electron beams:
\begin{eqnarray}
\sigma(e^+e^-)&=& 563 ~fb,\\ 
\sigma(e^+e^-_L)&=& 785 ~fb,\\ 
\sigma(e^+e^-_R)&=& 341 ~fb.
\end{eqnarray}

By using the results of $c'_{ii}$ in Table II. and $c_{ii}$ 
in Table III., we obtain the $1\sigma$ level statistcal 
errors of $g_i'$ and $g_i^{L,R}$ given in Table IV. From this table 
we see that the accuracies are about $10^{-18}~e~cm$ for $d_t^{\gamma,Z}$.  
In the unpolarized case, the best limit is on $Im(d_t^\gamma)$ which 
is the main component of $g_3'$. Better limits can be obtained by 
using polarized electron beams with the same integrated luminosity. 
In this case, one can only measure the combination of 
$d_t^\gamma$ and $d_t^Z$. One can combine the two modes of 
electron polarization to obtain $d_t^\gamma$ and $d_t^Z$ separately. But  
that needs two periods of running. Although with the right-handed electron 
beam, one gets a relative larger $c_{ii}$(cf. Table III.), the 
statistical errors  are the same as that with the left-handed 
electron beam. 

In conclusion, we have used the optimal observables to 
extract the CP-violating dipole couplings of the top quark 
at the NLC. The accuracies with which these couplings 
can be measured at  $\sqrt{s}=500$ GeV 
$e^+e^-$ collider with an integrated luminosity of $50fb^{-1}$ 
are about $10^{-18}~e~cm$.

\begin{table}
\caption{} 
Matrix elements $c_{ij}$ of $O_{2i}$ and $O_{4i}$ at $\sqrt{s}=500$ 
GeV unpolarized $e^+e^-$ collider.Unit: $10^4$ GeV$^2$.     
\small{
\begin{center}
\begin{tabular}{|cccccc|}
          & $O_{21}$ & $O_{22}$ & $O_{23}$ & $O_{24}$ &  \\ \hline
          & 1.85     & 1.40     & 0        & 0        & $O_{21}$ \\ 
          &          & 4.11     & 0        & 0        & $O_{22}$ \\ 
$O_{41}$  & 0.817    &          & 7.59     & 1.40     & $O_{23}$ \\ 
$O_{42}$  & 0.81     & 2.39     &          & 1.05     & $O_{24}$ \\
$O_{43}$  & 0        &  0       & 4.27     &          &          \\
$O_{44}$  & 0        &  0       & 0.80     & 0.46     &          \\\hline
          & $O_{41}$ & $O_{42}$ & $O_{43}$ & $O_{44}$ &          \\ 
\end{tabular}
\end{center}
}

\caption{} 
Matrix elements $A_{ij}$ and $c'_{ii}$ of $O_{2i}$ and $O_{4i}$ 
at $\sqrt{s}=500$ GeV unpolarized $e^+e^-$ collider.     
\small{
\begin{center}
\begin{tabular}{|c|cccc|c|}
&\multicolumn{4}{c}{$A_{ij}$} & $c_{ii}'(10^4$GeV$^2)$    \\ \hline
$O_{2i}$     &      0.90 & 0.43 & 0     & 0       & 1.18 \\ 
     &      $-0.43$ & 0.90    & 0     & 0       & 4.78 \\ 
     &       0   & 0       & 0.98  & $-0.20$ & 7.88 \\ 
     &       0   & 0       & 0.20  & 0.98    & 0.763 \\\hline 
$O_{4i}$     &       0.92 &   0.39   &   0 &   0   & 0.475 \\
     &      $-0.39$ & 0.92  &   0   &  0   & 2.73\\
     &      0       & 0  &  0.98    &  $-0.20$ & 4.43  \\
     &      0       & 0  &  0.20    & 0.98    &  0.299  \\ 
\end{tabular}
\end{center}
}
\caption{} 
Matrix elements $c_{ii}$ of $O_{2i}$ and $O_{4i}$ at 
$\sqrt{s}=500$ 
GeV $e^+e^-$ collider with left- and right- polarized electron beams. 
Unit: $10^4$ GeV$^2$.          
\small{
\begin{center}
\begin{tabular}{|c|cccc|}
          & $O_{21}$ & $O_{22}$ & $O_{41}$ & $O_{42}$   \\ \hline
$e^+e^-_L$ & 9.74  & 7.31 & 4.82 & 3.48 \\ 
$e^+e^-_R$ & 22.2  & 17.4 & 10.9 & 8.23\\\hline
\end{tabular}
\end{center}
}

\caption{} 
$1\sigma$ statistical errors of the coupling constants $g_i'$ 
and $g_i^{L,R}$ at $\sqrt{s}=500$ GeV collider 
with an integrated luminosity $50fb^{-1}$.Unit:$10^{-18}~cm$.   
\small{
\begin{center}
\begin{tabular}{|c|cccccccc|}
& $g_{1}'$ & $g_{2}'$ & $g_{3}'$ & $g_{4}'$ 
& $g_1^L$ & $g_2^L$ & $g_1^R$ & $g_2^R$   \\ \hline
$O_{2i}$ &6.35 & 3.15 & 2.46 & 7.90 & 1.87 & 2.16 & 1.88 & 2.12\\
$O_{4i}$ &7.94 & 3.31 & 2.60 & 10.0 & 2.11 & 2.48 & 2.13 & 2.45\\
combined &4.96 & 2.28 & 1.79 & 6.20 & 1.40 & 1.63 & 1.41 & 1.60\\ 
\end{tabular}
\end{center}
}

\end{table}

\begin{center}
{\bf Acknowledgements} 
\end{center}

The author is financially supported by the Alexander von Humboldt  
Foundation of Germany.

\end{document}